\documentclass[11pt]{article}

\usepackage{amsmath}
\usepackage{graphicx}
\usepackage{enumerate}
\usepackage{natbib}
\usepackage{typearea}
\usepackage[hyphens]{url}
\usepackage{hyperref}
\usepackage{caption}
\usepackage{subcaption}
\usepackage{xcolor}

\hypersetup{
           breaklinks=true,   
           hidelinks,  
           pdfusetitle=true,  
        }

\usepackage{amsfonts, amsmath, amssymb, amsthm, dsfont}
\usepackage{amsthm}
\usepackage{natbib}
\usepackage{booktabs}
\usepackage{multirow}

\usepackage{bbm,bm}

\def\inL2{\,{\buildrel L_2 \over \rightarrow}\,}

\def\pp{\partial}

\newcommand{\m}[1]{\begin{pmatrix}
#1
\end{pmatrix}}

\newcommand{\cur}[1]{\left\{#1\right\}}
\newcommand{\pa}[1]{\left(#1\right)}

\def\ee{\ensuremath{\mathrm{e}}}

\def\dd{\ensuremath{\mathrm{d}}}

\title{Mortality Forecasting using Variational Inference}
\author{
Patrik Andersson\thanks{Uppsala University ({patrik.andersson@statistics.uu.se})}
\and  
Mathias Lindholm\thanks{Stockholm University ({lindholm@math.su.se})}
} 

\date{}

\begin{document}

\maketitle

\begin{abstract}
This paper considers the problem of forecasting mortality rates. A large number of models have already been proposed for this task, but they generally have the disadvantage of either estimating the model in a two-step process, possibly losing efficiency, or relying on methods that are cumbersome for the practitioner to use. We instead propose using variational inference and the probabilistic programming library Pyro for estimating the model. This allows for flexibility in modelling assumptions while still being able to estimate the full model in one step.

The models are fitted on Swedish mortality data and we find that the in-sample fit is good and that the forecasting performance is better than other popular models.

Code is available at \url{https://github.com/LPAndersson/VImortality}. \\\\
\textbf{Keywords:} Non-linear state-space-models; Mortality forecasting; Hidden Markov model, Variational inference
\end{abstract}

\section{Introduction}
Attempts to forecast mortality go back at least as far as \cite{gompertz1825nature}. A more recent example is the Lee-Carter model \citep{lee1992modeling} and its extensions, see \cite{booth2008mortality, haberman2011comparative, carfora2017quantitative} for a survey. Applications of mortality forecasting can be found in for example demographic predictions and in the insurance industry.

The Lee-Carter model is a log-linear multivariate Gaussian model of mortality rates. A major criticism of Lee-Carter-type models is that the model training is done as a two-step process. In the first step, point estimates of the mortality rates are obtained, for example as the maximum likelihood estimate of a Poisson distribution, and in the second step, a latent process is fitted to these estimates. This method has the advantage that it is simple and fast to implement, but it is inefficient when compared to the simultaneous estimation of all unknown parameters. Also, it is not possible to distinguish between the finite population noise of the mortality estimates and the noise from the latent process. Both of these issues can potentially affect the quality of the forecasts. 

Simultaneous estimation of parameters has been considered in \cite{andersson2020mortality} where particle filtering methods are used to estimate a state-space model with Poisson distributed observations, similar to \cite{brouhns2002poisson}. However, this method has its drawbacks. It could be considered cumbersome for practitioners as it requires custom implementation and tuning, and since the particle filter methods are computationally expensive, the number of parameters must not be too large. The complexity of the modelling is reduced when changing the observational model from a Poisson distribution to a Gaussian, see e.g.\ \cite{de2006extending} for a state-space model treatment of a Gaussian Lee-Carter model.

Recently it has been suggested to use models from deep learning, sometimes called deep factor models, to forecast high-dimensional multivariate time series. Some examples of this can be found in \cite{nguyen2021temporal, wang2019deep, salinas2020deepar, rangapuram2018deep}. The applications presented in those articles differ from mortality forecasting in the scale of the problem. In mortality forecasting, the dimension of the time series is about 100 (the lifetime in years of a human) and the number of observations of the time series is also about 100 (although some countries do have reliable data for longer than that). As a consequence, to avoid overfitting, we need to consider simpler models. This includes simpler functions for mapping latent variables to the observed time series, linear Gaussian models instead of RNNs for propagating the latent variables forward in time and fewer latent factors.

Compared to previous mortality forecasting models, the novelty of this paper is therefore to use black-box variational inference \citep{ranganath2014black} to solve the inference problem. This means that after specifying how to sample from the model and the approximate posterior, the inference is done automatically without any model-specific customisation. The family of models that can be handled is also expanded. For example, one can consider other families of functions for mapping the latent process to mortality rates. Also, in this case, all the parameters can be estimated simultaneously. This latter point is problematic when using particle filter techniques, where it is necessary to estimate the linear mapping from the latent process to the mortality rates first, before continuing with the estimation of the other parameters.

Other approaches that use machine learning techniques for forecasting mortality rates can be found in e.g.\ \cite{richman2019lee,richman2021neural,perla2021time} that consider various types of Gaussian recurrent neural network structures, \cite{nigri2019deep,marino2020measuring,lindholm2022efficient} that consider univariate LSTM neural network, both with and without a Poisson population assumption, and \cite{deprez2017machine} that consider tree-based techniques.

The model is implemented using the probabilistic programming language Pyro \citep{bingham2018pyro} and the code is available at \url{https://github.com/LPAndersson/VImortality}.

The rest of the paper is organised as follows: In Section \ref{sec:model} we describe the probabilistic model that will be used for forecasting. In Section \ref{sec:vi} we give a brief introduction to variational inference and in Section \ref{sec:forecastValidation} we describe how to forecast the mortality once the model has been trained and how we validate the forecast. In Section \ref{sec:results} we demonstrate our method on an example and compare to other models. Section \ref{sec:conclusions} concludes the paper.

\section{Model}\label{sec:model}
In this section, we describe the probabilistic model that defines the mortality dynamics. Uppercase letters will denote random variables and lowercase letters the corresponding observed value. Greek letters will denote unknown parameters that are to be estimated. 

Mortality data can be aggregated in different forms and clearly the choice of model will have to be adjusted accordingly. For example, the data could contain the population size of each age group at the beginning of the year and the number of deaths during the year. In this case, a binomial model seems natural. We however consider data on the yearly number of deaths and the exposure to risk in each age group. The exposure to risk in this setting is the total time that the individuals in the population were a certain age in a certain year.  The number of deaths in age $a\in \cur{0,1,\ldots, \bar a}$, year $t\in \cur{0,1,\ldots, \bar t}$ is denoted by $D_{a,t}$ and the exposure by $E_{a,t}$. 

Our model is a state-space model that can be written as:
\begin{align}
    & D_{a,t}\mid X_t,E_{a,t} \sim  \mathsf{Poisson}\pa{E_{a,t}\exp\pa{f^\psi_{a}(X_t)}},\label{eq:deathDist}\\
    & X_{i,t} =X_{i,t-1} + K_{i,t-1} + U_{i,t},\quad U_{i,t} \text{ iid }  \mathsf{N}(0,\sigma^2_{X,i}),\label{eq:state_level}\\
    & K_{i,t} = \mu_i + \varphi_i(K_{i,t-1} - \mu_i) + V_{i,t},\quad V_{i,t} \text{ iid }   \mathsf{N}(0,\sigma^2_{K,i}).\label{eq:state_trend}
\end{align}
Here $i=1,2,\ldots , d$, where $d$ is the dimension of the latent variables. We also require $0\leq \varphi_i \leq 1$. The function $f^\psi_a$ is the $a$:th component of $f^\psi:\mathbb R^d \to \mathbb R^{\bar a+1}$. In our examples in Section \ref{sec:results}, $f^\psi$ will be given by either an affine transformation or a sum of radial basis functions. That is either, $f^\psi(x) = Ax + b$, where $A$ and $b$ are trainable or the $a$th component of $f^\psi$ is
$$
f^\psi_a(x) = x^T\sum_{i=1}^p w_i \ee^{-\tau^2(\frac{a-\mu_i}{\bar a +1})^2} + b_a.
$$
We will fix $\tau^2$ and therefore $\cur{w_i ,\mu_i}_{i=1}^p$ and $\cur{b_a}_{a=0}^{\bar a}$ are the trainable parameters. Compared to the more general affine transformation, radial basis functions have the advantage of inducing a certain smoothness of $f_a$ as a function of $a$, encoding a prior that similar ages should have similar mortality.

We remark here that the exact specification of the above model is not critical for the continuation. For example, the exponential link function in the Poisson distribution could be changed to some other positive differentiable function without complication. We are assuming that the components of the latent process are independent, instead, we let any dependence be captured by $f$. However, this latent process could be replaced with some other Markov process.

\section{Variational inference}\label{sec:vi}
Here we explain the main ideas of variational inference in a general setting. At the end of the section, we connect this to our specific model. For more on variational inference in general we refer to \cite{ranganath2014black} and for the application to state space models, see \cite{archer2015black}.
    
We are observing $y$, whose distribution depends on a latent variable $x$ and an unknown parameter $\psi$. This is modelled by the joint distribution
$$
p_\psi(y,x) = p_\psi(y\mid x)p_\psi(x).
$$
The likelihood,
$$
    L(\psi)=p_\psi(y) = \int p_\psi(y,x)\dd x,
$$
is in general not tractable and therefore approximations are needed in order to be able to estimate $\psi$.
Consider a parametrised distribution, the approximate posterior, $q_\theta(x)$. Then observe that, due to Jensen's inequality, the log-likelihood is
\begin{align*}
    l(\psi):=&\log L(\psi) = \log \int \frac{p_\psi(y,x)}{q_\theta(x)} q_\theta(x)\dd x  \\
    \geq &\int \pa{\log p_\psi(y,x) - \log q_\theta(x)}q_\theta(x)\dd x=:\mathcal L(\psi,\theta).
\end{align*}
The right-hand side is known as the evidence lower bound (ELBO). The idea of variational inference is to instead of maximising the log-likelihood, maximise the ELBO. Towards this, we calculate the gradients
\begin{align*}
    \partial_\psi \mathcal L(\varphi,\theta) &= \int \partial_\psi \log p_\psi(y,x)q_\theta(x) \dd x ,\\
    \partial_\theta \mathcal L(\psi,\theta) &=  \int (\log p_\psi(y,x) - \log q_\theta(x))\partial_\theta \log q_\theta(x)  q_\theta(x)   \dd x.
\end{align*}

We can then proceed to obtain unbiased estimates of the gradients by sampling from $q_\theta$ and maximise $\mathcal L$ using stochastic optimisation algorithms. Once converged,  $q_\theta(x)$ can be used as an approximation of the posterior distribution of the latent variables $p_\psi(x\mid y)$.

Further, to obtain faster convergence, various variance reduction techniques are often used. Here we only mention the so-called reparametrisation trick. Suppose that we can find functions $x_\theta$ such that
\begin{equation}\label{eq: reparametrisation trick}
    \int f(x) q_\theta(x)\dd x = \int f(x_\theta(z))q(z)\dd z,
\end{equation}
which makes the sampling distribution independent of $\theta$. In particular, the gradient satisfies
$$
    \partial_\theta \int f(x) q_\theta(x)\dd x = \partial_\theta \int f(x_\theta(z))q(z)\dd z = \int \pp_\theta f(x_\theta(z))q(z)\dd z,
$$
which usually improves the sampling variance, compared to differentiating the density directly. An important example of a distribution that allows for reparametrisation according to \eqref{eq: reparametrisation trick} is the Gaussian, since if $Z\sim \mathsf N(0,1)$ then $\mu + \sigma Z\sim \mathsf N (\mu,\sigma^2)$.

In the numerical illustrations in Section~\ref{sec:results}, the approximate posterior is modelled as a Gaussian distribution with an autoregressive covariance. That is, the distribution of the process is given by
\begin{align*}
    \tilde X_{i,t} &= \tilde\mu^X_{i,t } + \alpha_{i,t}\tilde X_{i,t-1}  + \tilde e^X_{i,t},\quad \tilde e^X_{i,t} \text{ iid } \mathsf N(0,\tilde\sigma^2_{X,i}),\\
    \tilde K_{i,t} &= \tilde \mu^K_{i,t} + \beta_{i,t}\tilde K_{i,t-1} + \rho_{i,t}\tilde X_{t-1} + \tilde e^K_{i,t},\quad \tilde e^K_{i,t} \text{ iid } \mathsf N(0,\tilde \sigma^2_{K,i}).
\end{align*}

\section{Forecasting and validation}\label{sec:forecastValidation}
By maximizing the ELBO we have obtained estimates of $\varphi$ and $\theta$ and the joint distribution of $(\tilde X_{i,\bar t},\tilde K_{i,\bar t})$. This allows us to proceed with forecasting the mortality, as will be discussed in this section.

Since both the approximate posterior and the latent process is Gaussian, the forecasting distribution of the latent process is also Gaussian. That is, for $t>\bar t$,
$$
\m{
\hat X_{i,t}\\ \hat K_{i,t}} 
\sim \mathsf N\pa{
\m{\hat\mu^X_{i,t}\\
\hat\mu^K_{i,t}}
,
\m{
\hat \sigma^2_{X,t} & \hat\rho_{i,t}\hat \sigma_{X,t}\hat \sigma_{K,t}\\
    \hat\rho_{i,t}\hat \sigma_{X,t}\hat \sigma_{K,t} & \hat \sigma^2_{K,t}
}
},
$$
where the parameters can be calculated iteratively from \eqref{eq:state_level} and \eqref{eq:state_trend}, by using the initial value
$$
    \m{
\hat X_{i,\bar t}\\ \hat K_{i,\bar t}}:=  	\m{
\tilde X_{i,\bar t}\\ \tilde K_{i,\bar t}},
$$
and  the forecast of mortality rates is given by $\exp(f^{\hat \psi}(\hat X_t))$.

If one wants to forecast the actual number of deaths, a forecast of the number of living at the beginning of the year is also needed, together with some assumption on the distribution of when in the year people are born. For a longer discussion on how this can be done, we refer to \cite{andersson2020mortality}.

The forecast is validated by calculating the logarithmic score of the forecast on the validation data set. The logarithmic score is the logarithm of the predicted density evaluated at the observed value, see for example \cite{gneiting2007strictly}. That is, if $\mathcal P_{a,t}$ is the forecasted distribution of death counts from \eqref{eq:deathDist} and $d_{a,t}$ is the observation, the log-score is
$$
\text{logs}(\mathcal P_{t}, d_{t}) := \sum_a \log \mathcal P_{a,t}(d_{a,t}).
$$
We will evaluate the models using a rolling window of training and evaluation data and calculate the average score over these windows, which we call $\overline{ \text{logs}(\mathcal P, d) }$. An alternative that we also calculate is
$$
R^2 = \frac{\overline{ \text{logs}(\mathcal P, d) } - \overline{ \text{logs}(\overline{\mathcal P}, d) }}{\overline{ \text{logs}(\widehat{\mathcal P}, d) } - \overline{ \text{logs}(\overline{\mathcal P}, d) }}.
$$
Here, $\overline{\mathcal P}$ denotes the model with only a constant intercept and $\widehat{\mathcal P}$ the saturated model, i.e.\ the model that forecasts with mean equal to the observation.

\section{Results}\label{sec:results}

In this section, we illustrate the performance of the models by fitting it to a dataset on the mortality of Swedish males. The dataset is collected from \cite{hmd2018data}. We evaluate the models using a rolling window; using 60 years to fit the model and 10 years to evaluate the forecast against the actual outcome. The training and evaluation windows are then rolled forward one year and the process repeats.

The model is as in \eqref{eq:deathDist} - \eqref{eq:state_trend} where $f$ is either affine or a sum of radial basis functions. We begin by selecting hyperparameters for each model, e.g.\ dimension of the affine transformation or the number of radial bases. We then illustrate the fitted model and the out-of-sample forecasting performance. Finally, we compare our model with the Lee-Carter model and the model by \cite{plat2009stochastic}.

\subsection{Affine}\label{seq:results_sweden}

\begin{figure}
    \centering
    \begin{subfigure}[t]{0.49\textwidth}
        \centering
        \includegraphics[width=\textwidth]{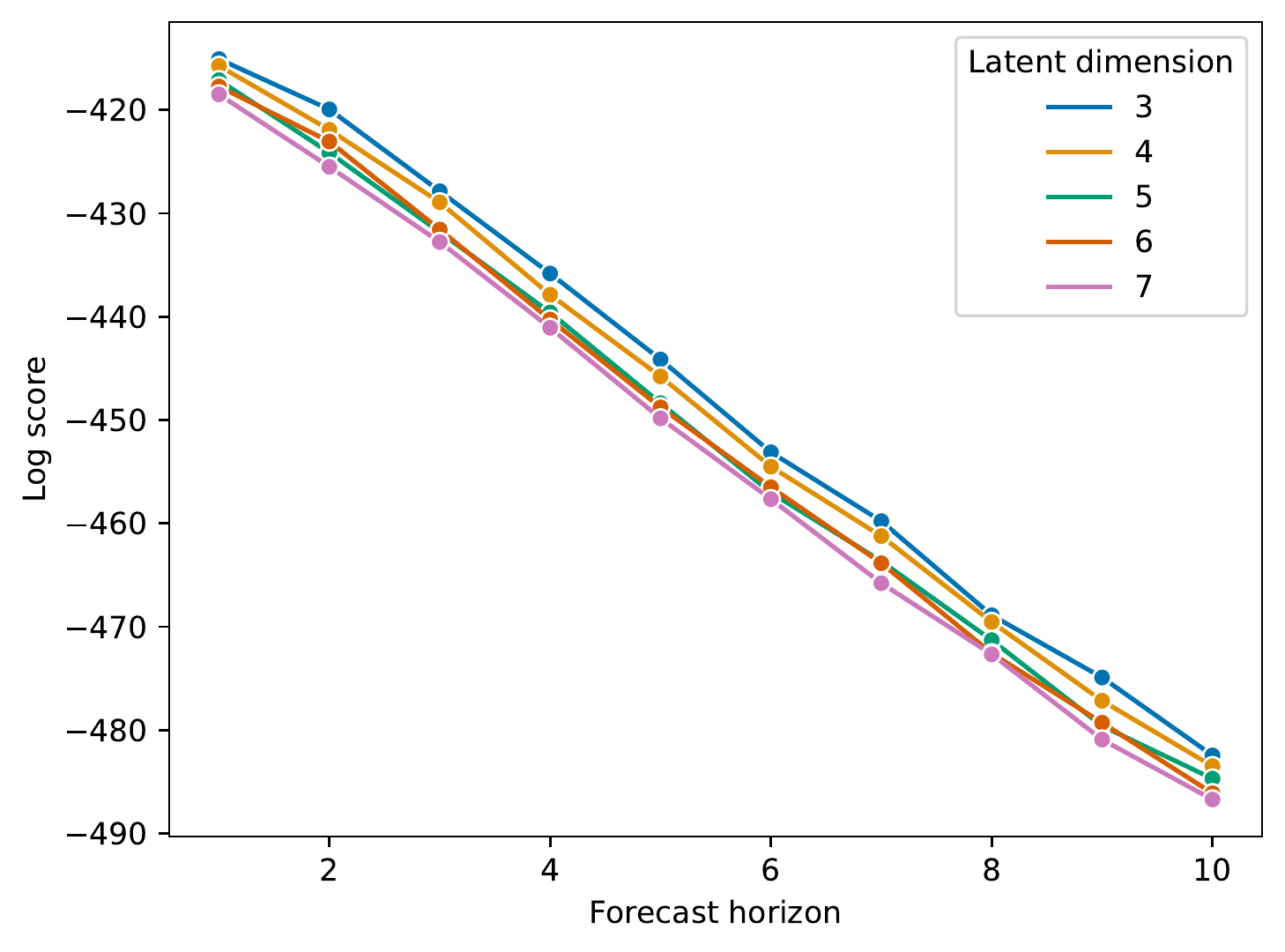}
        \caption{}
        \label{fig:model_evaluation_sweden_logscore}
    \end{subfigure}
    \hfill
    \begin{subfigure}[t]{0.49\textwidth}
        \centering
        \includegraphics[width=\textwidth]{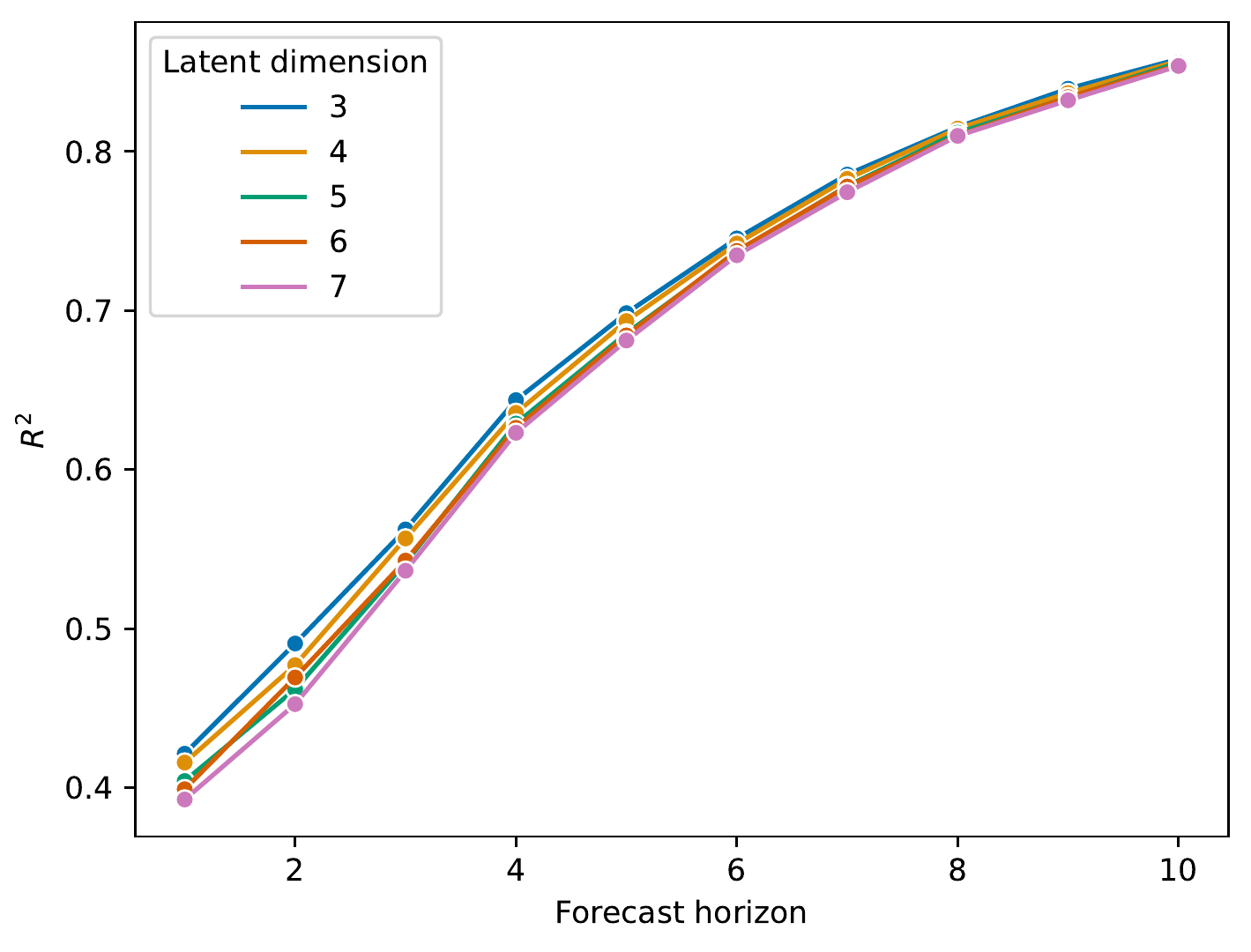}
        \caption{}
        \label{fig:model_evaluation_sweden_R2}
    \end{subfigure}
    \hfill
        \caption{Affine model: Evaluation of the model when fitted on Swedish male mortality data with the first year from 1931 to 1952. Each model is fitted using 60 years of data and evaluated on the following 10 years. The figures show that 3 latent dimensions give the best out-of-sample performance.}
        \label{fig:affine_model_evaluation_sweden}
\end{figure}

\begin{figure}
    \centering \begin{subfigure}[t]{0.45\textwidth}
        \centering
        \includegraphics[width=\textwidth]{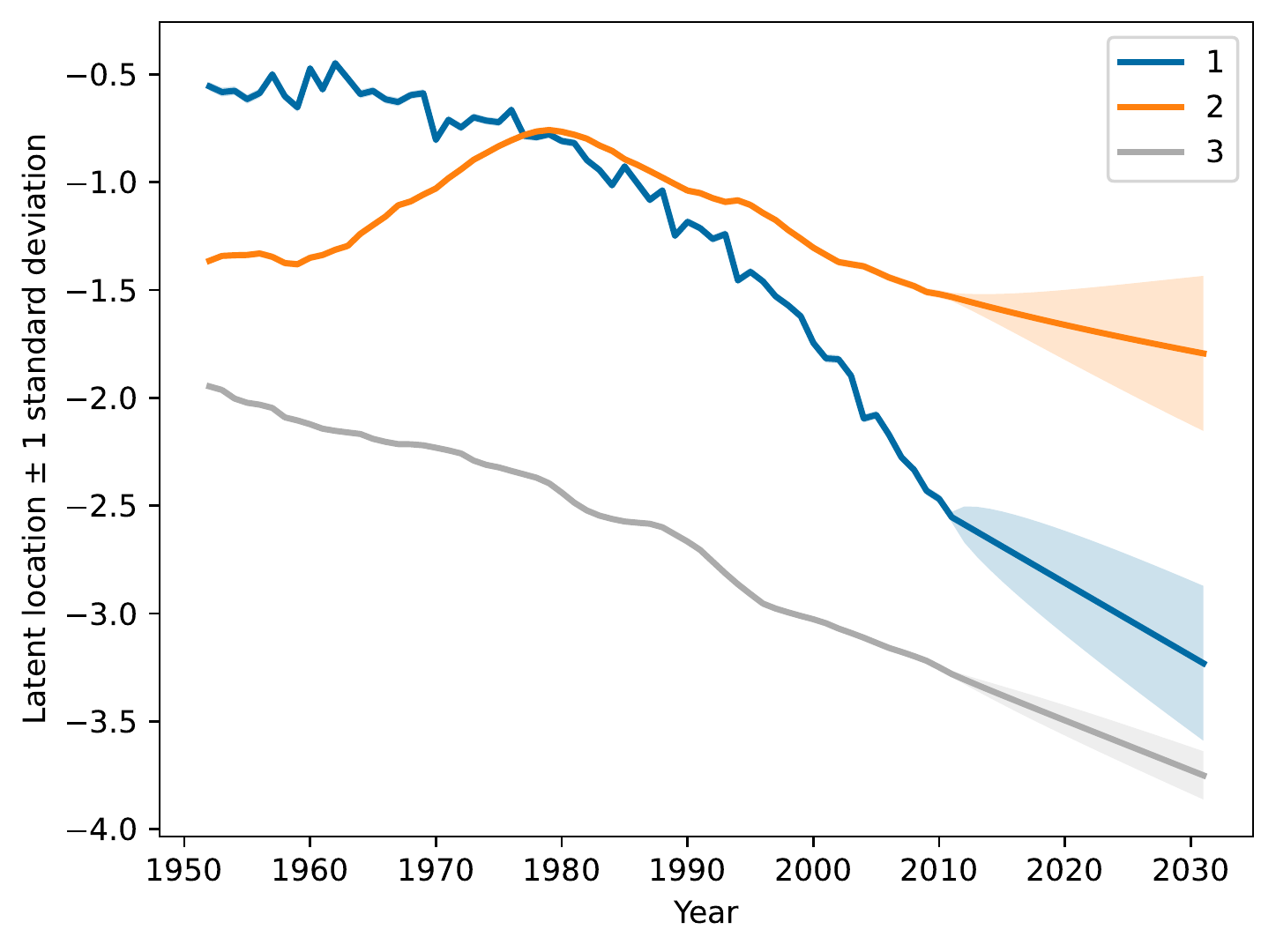}
        \caption{Level of latent process in-sample and its forecast. Shaded region is $\pm$ 1 standard deviation. From 1952 to 2011 shows smoothed values, from 2012 shows forecast.}
    \end{subfigure}
    \hfill
    \begin{subfigure}[t]{0.45\textwidth}
        \centering
        \includegraphics[width=\textwidth]{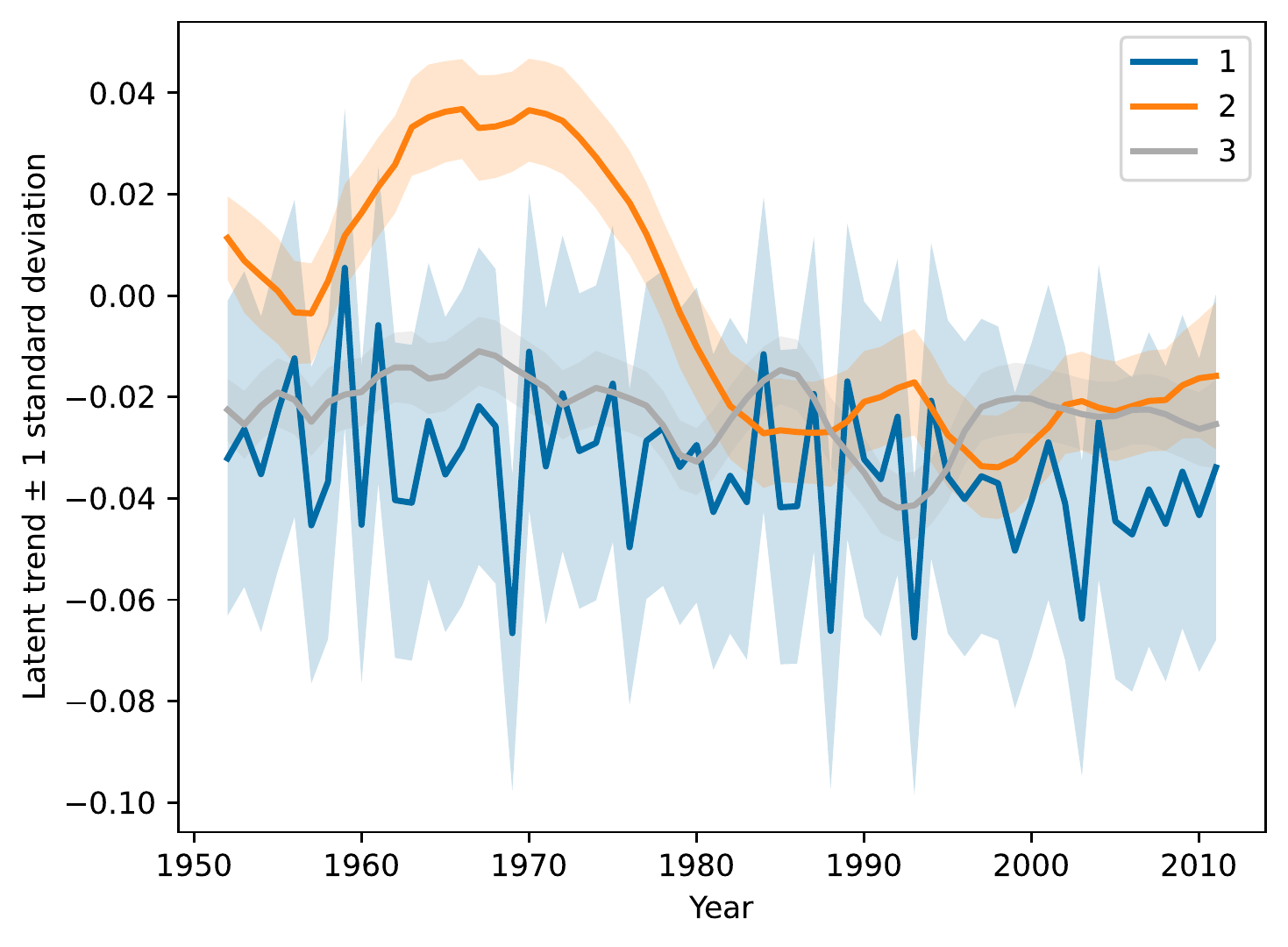}
        \caption{Trend of latent process. The shaded region is $\pm$ 1 standard deviation.}
    \end{subfigure}
    \hfill
    \begin{subfigure}[t]{0.45\textwidth}
        \centering
        \includegraphics[width=\textwidth]{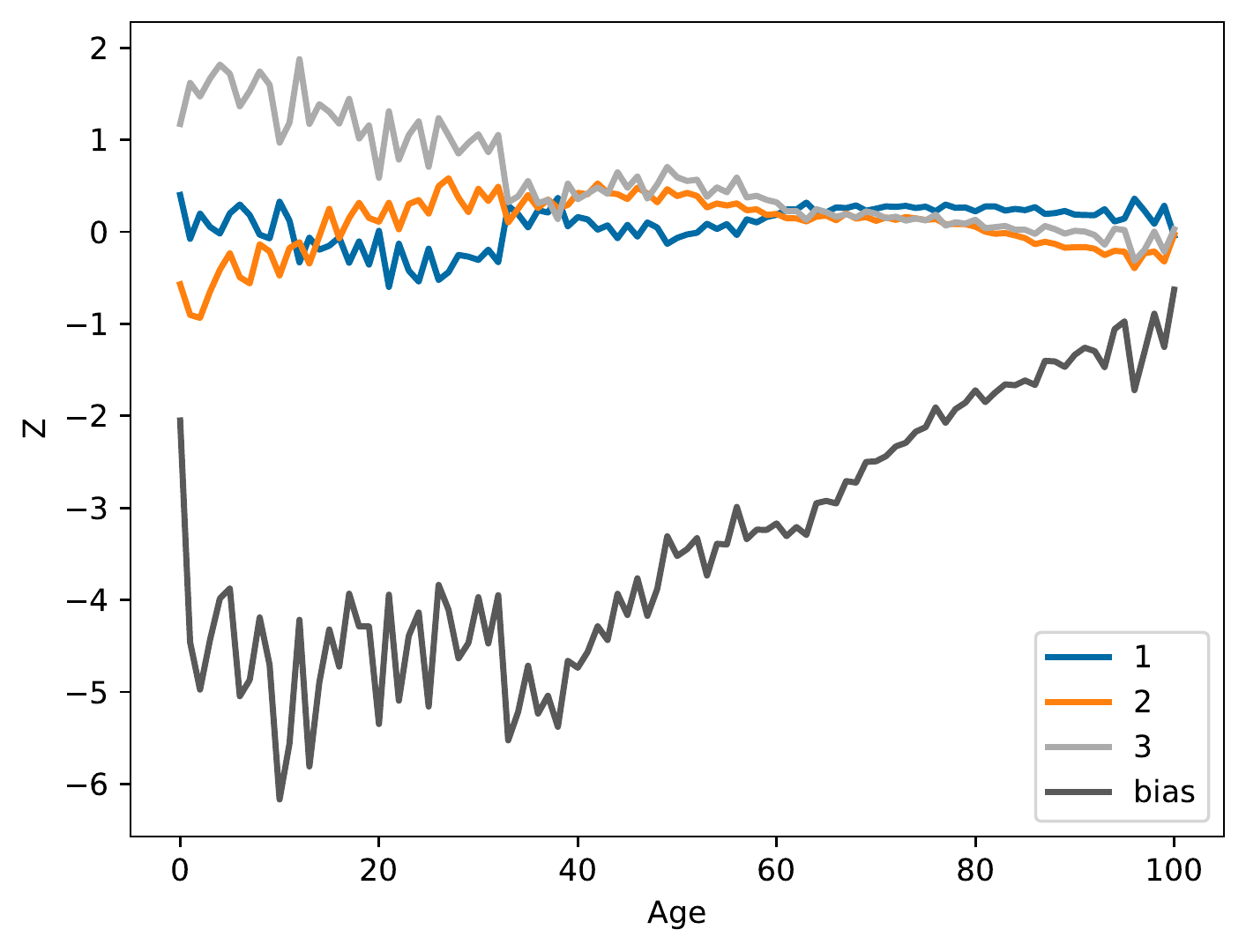}
        \caption{Factor loadings for the model with 3 latent dimensions}
    \end{subfigure}
    \hfill
    \begin{subfigure}[t]{0.45\textwidth}
        \centering
        \includegraphics[width=\textwidth]{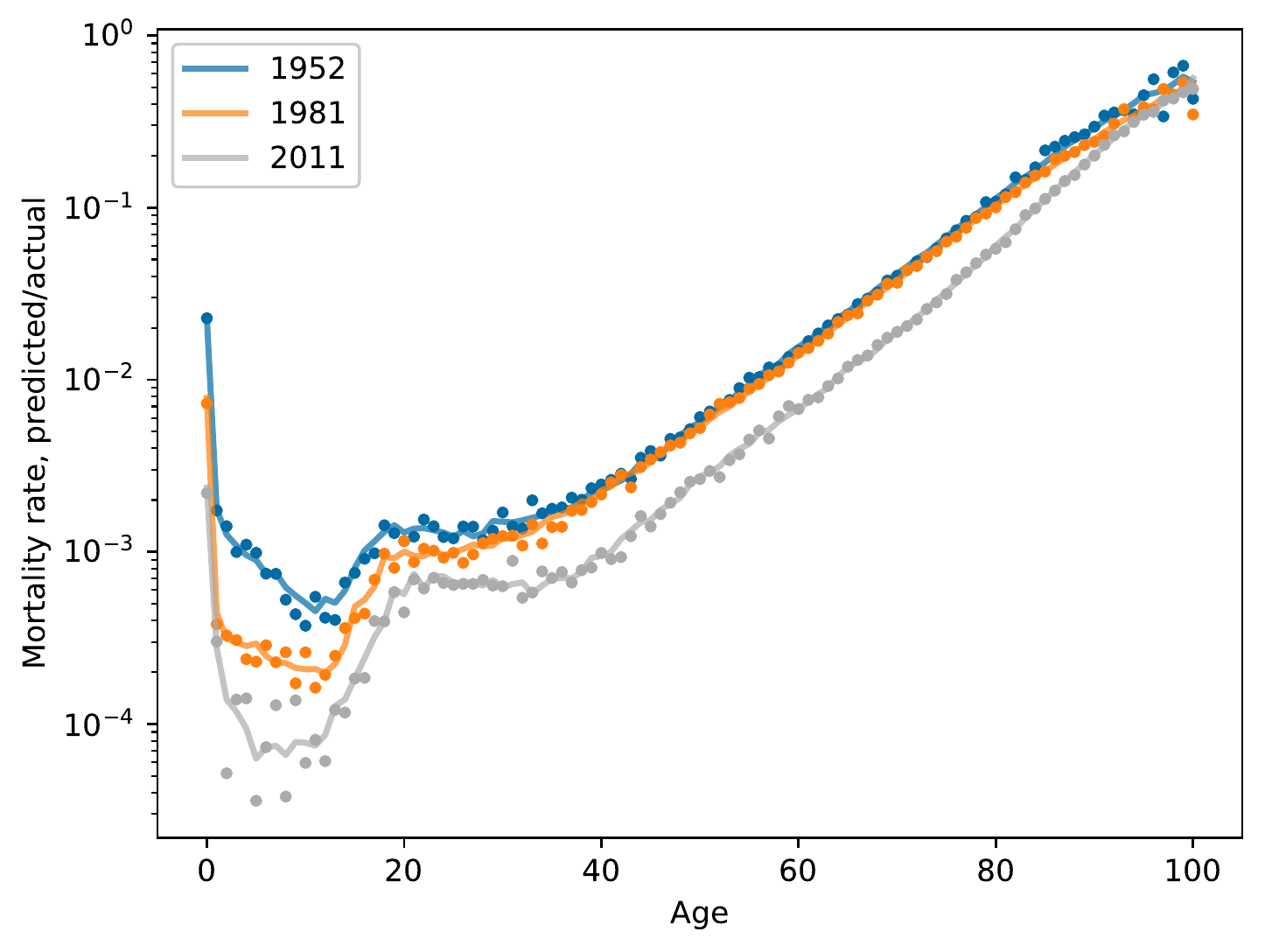}
        \caption{In-sample fit of model mortality rates. Dots are observed rates and solid lines are the fitted model. This model is fitted on data from 1952 to 2011.}
        \label{fig:linear_mortality_rates_in_sample_fit_sweden}
    \end{subfigure}
    
    \caption{Affine model: Illustration of the model fitted to Swedish data from 1952 to 2011 with three latent dimensions.}
    \label{fig:linear_model_illustration_sweden}
\end{figure}

\begin{figure}
    \centering
    \includegraphics[width=0.5\textwidth]{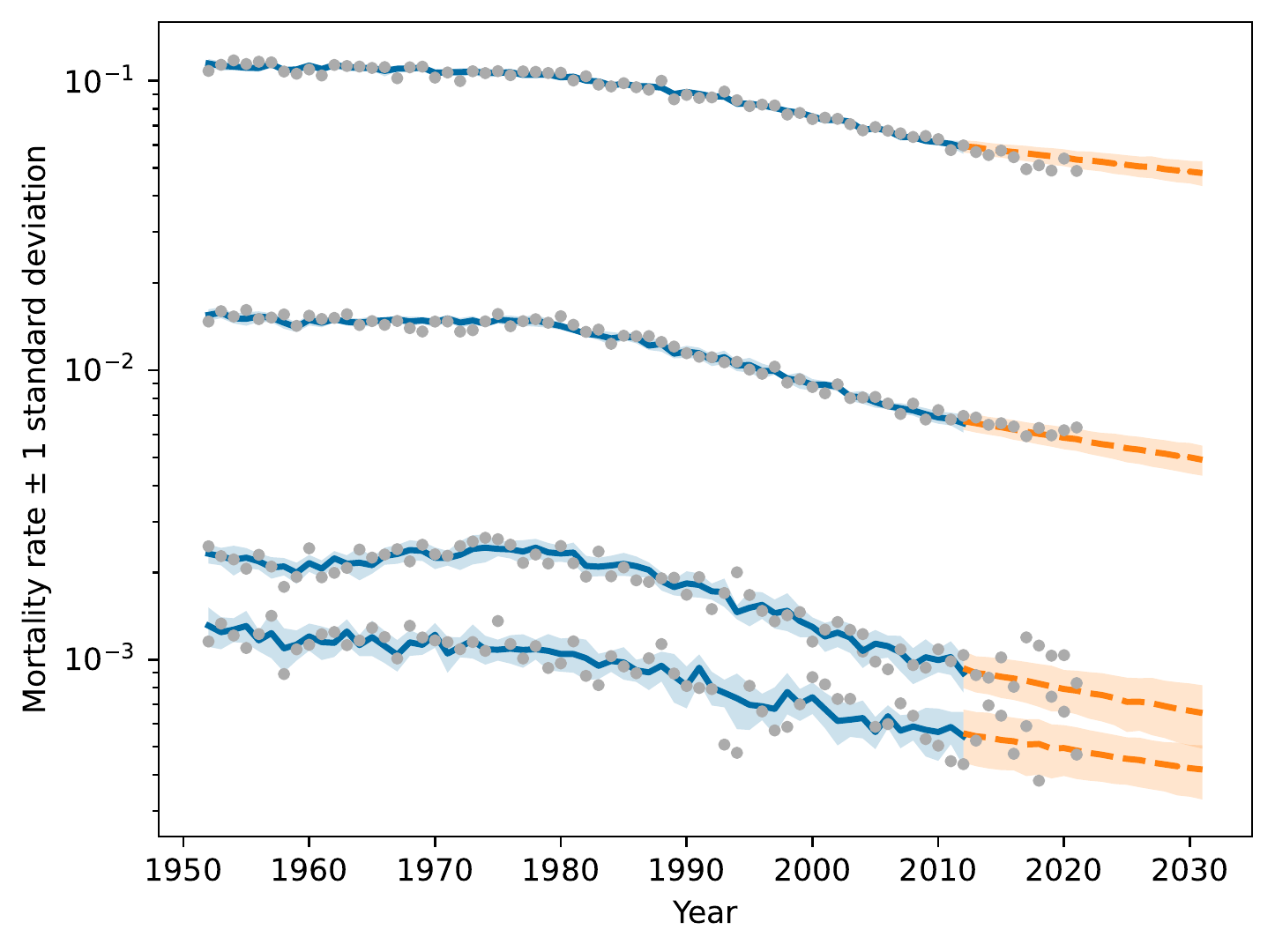}
    \caption{Affine model: Mortality forecasts for the age groups 20, 40, 60 and 80 (from the bottom up). The solid line is the smoothed mortality rate, the dashed line is the mean forecasted mortality rate and the shaded region is $\pm$ 1 standard deviation of the forecasted realised mortality rate. Dots indicate observed mortality rates.}\label{fig:model_linear_forecast_sweden}
\end{figure}

For model selection, we compare the out-of-sample log-score for varying dimensions of the latent process. In Figure \ref{fig:affine_model_evaluation_sweden} we see that 3 latent dimensions overall performs best. We have performed experiments also for dimensions 1 and 2, but the performance was considerably worse, and they are therefore excluded from the figure.
Figure \ref{fig:linear_model_illustration_sweden} illustrates the fitted model. In particular, we note in Figure \ref{fig:linear_mortality_rates_in_sample_fit_sweden} that the in-sample fit of the mortality rates  are quite good.
Figure \ref{fig:model_linear_forecast_sweden} shows the smoothed and forecasted mortality rates for the ages 20, 40, 60 and 80. The shaded regions represent $\pm$ 1 standard deviation and the grey dots are the observed mortality rates. That is, we should expect that around 7 out of 10 observations are within the shaded region. The pictures seem to confirm this.  

\subsection{Radial basis functions}

This section follows the same pattern as the previous one. We choose hyperparameters for the radial basis functions and illustrate the fitted model. In all models we choose $\tau=10$. Although this is a parameter that could be trained, our experiments show that it is difficult to train well. Our choice of $\tau$ corresponds to a typical width of the radial basis of about 14 years, which seems reasonable.

\begin{figure}
    \centering
    \begin{subfigure}[t]{0.49\textwidth}
        \centering
        \includegraphics[width=\textwidth]{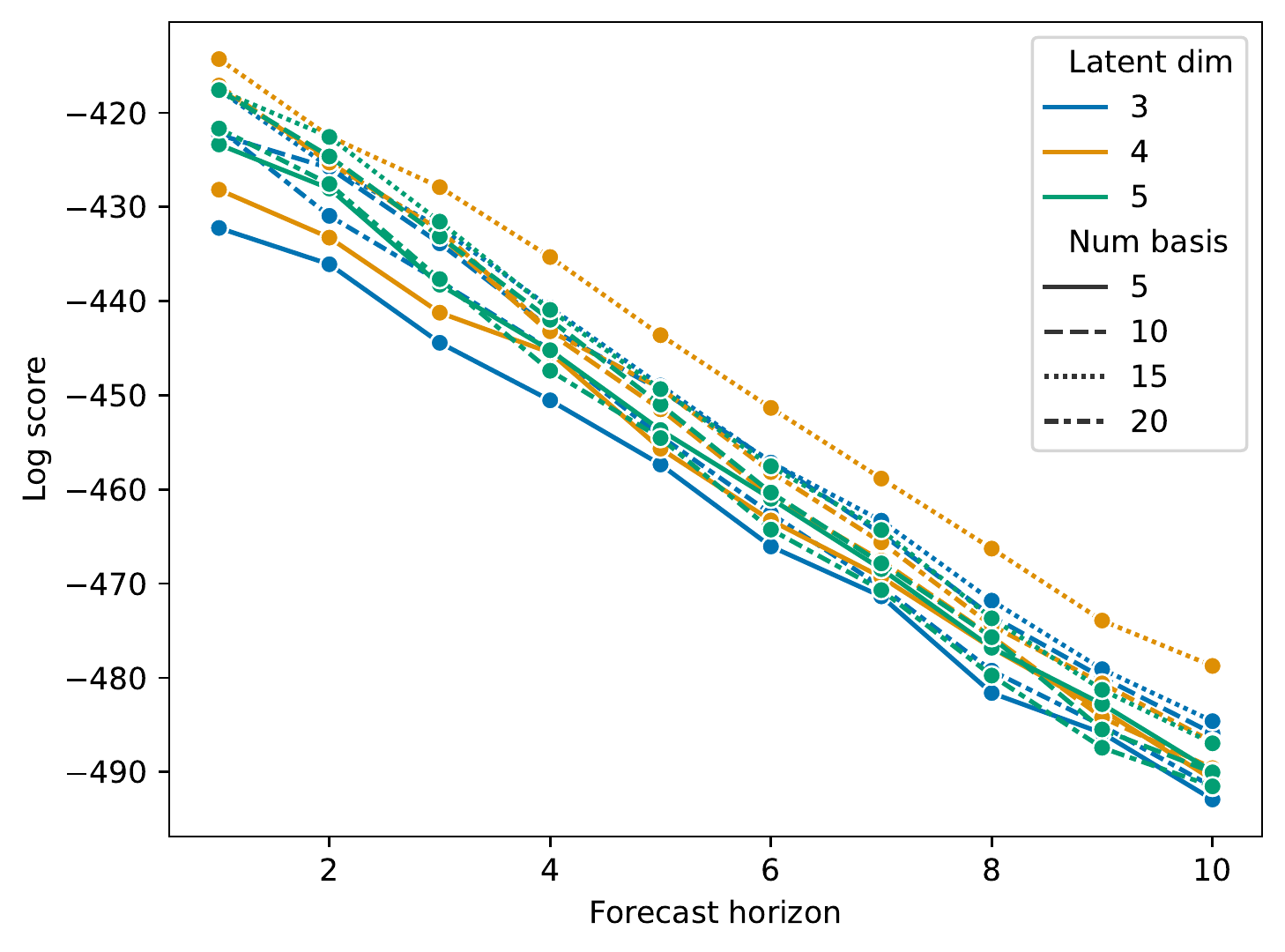}
        \caption{}
    \end{subfigure}
    \hfill
    \begin{subfigure}[t]{0.49\textwidth}
        \centering
        \includegraphics[width=\textwidth]{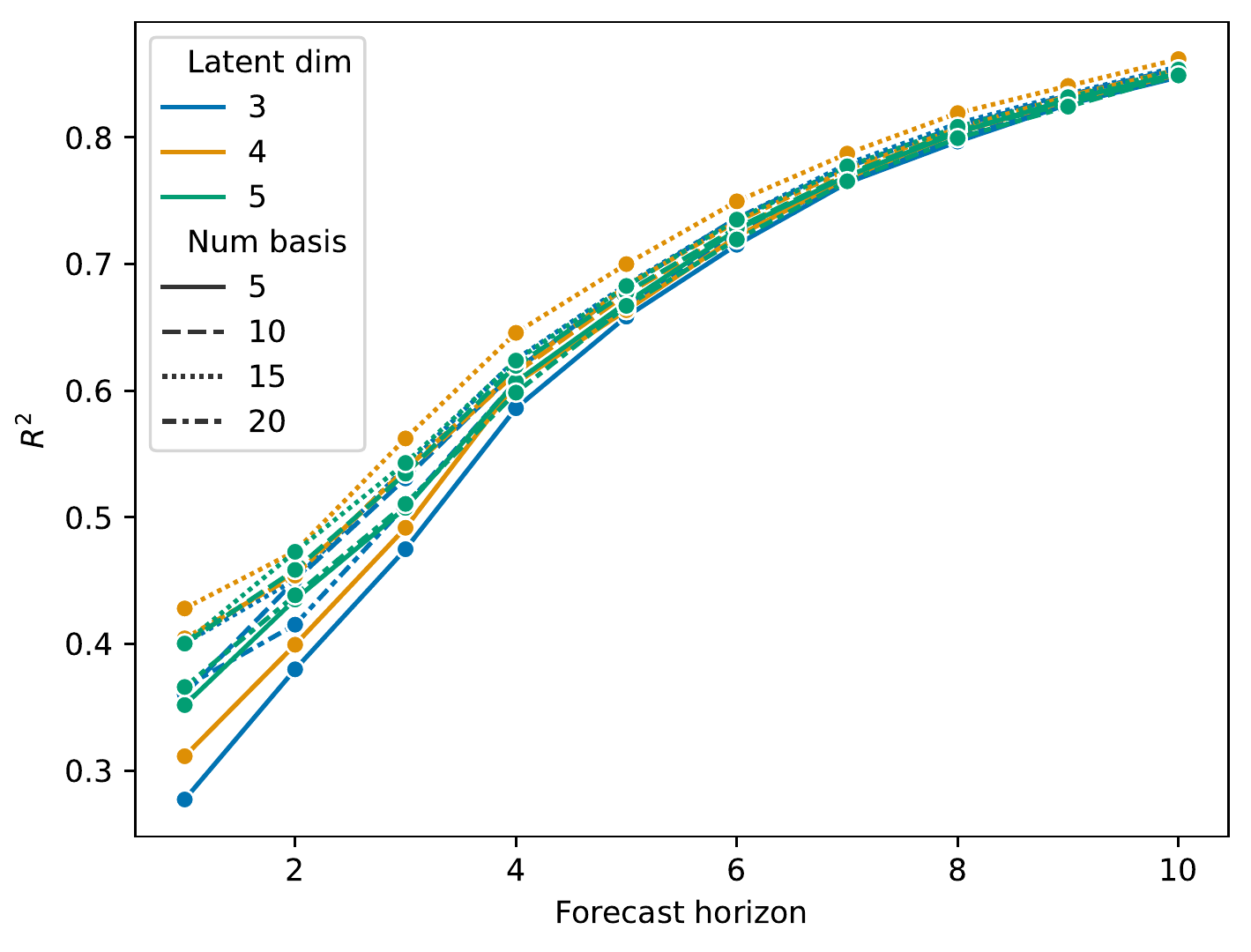}
        \caption{}
    \end{subfigure}
    \hfill
        \caption{Radial basis model: Evaluation of the model when fitted on Swedish male mortality data with the first year from 1931 to 1952. Each model is fitted using 60 years of data and evaluated on the following 10 years. The figures show that 4 latent dimensions and 15 radial basis functions give the best out-of-sample performance.}
        \label{fig:radial_model_evaluation_sweden}
\end{figure}

\begin{figure}
    \centering
    \begin{subfigure}[t]{0.45\textwidth}
        \centering
        \includegraphics[width=\textwidth]{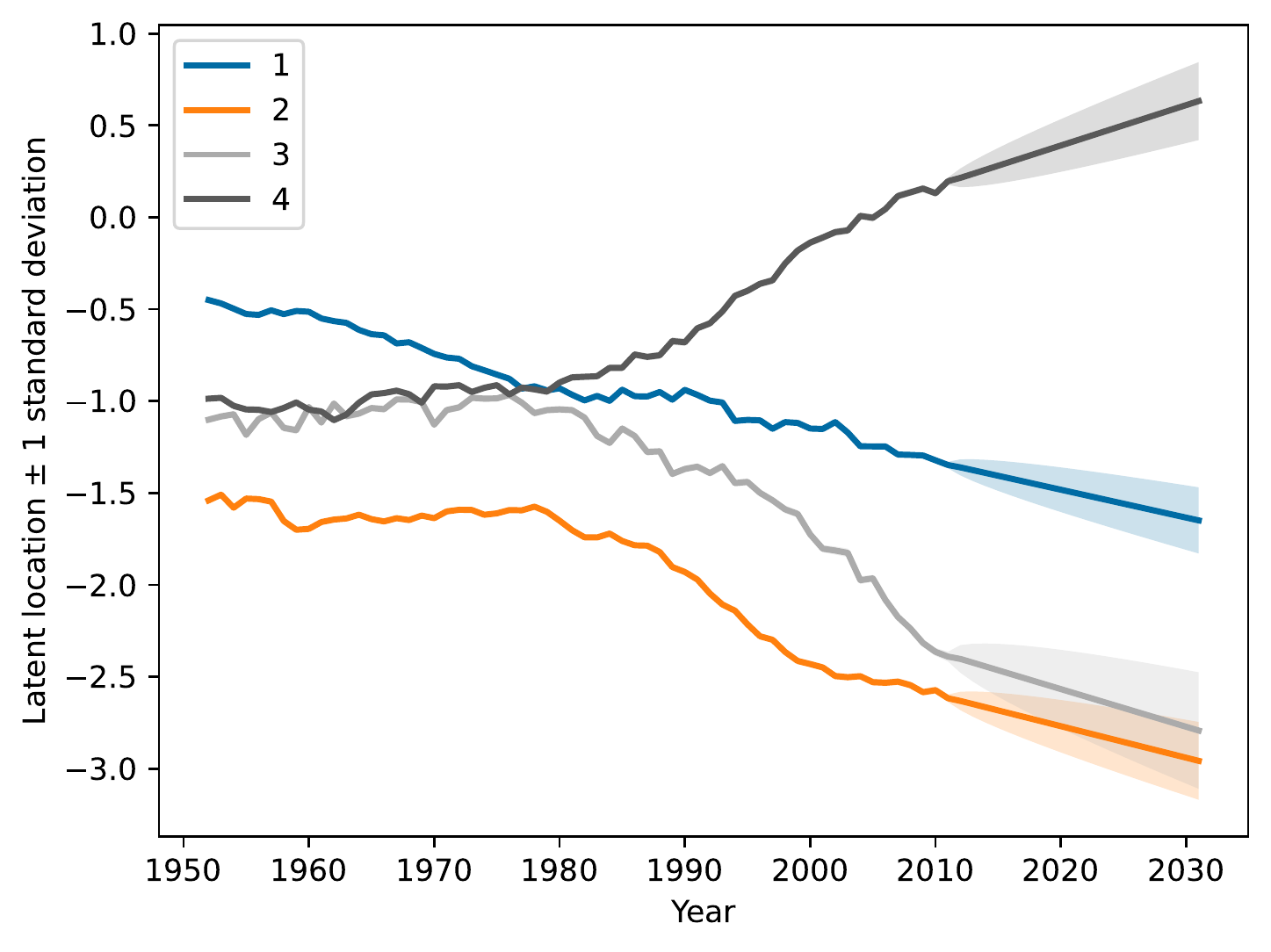}
        \caption{Level of latent process in-sample and its forecast. Shaded region is $\pm$ 1 standard deviation. From 1952 to 2011 shows smoothed values, from 2012 shows forecast.}
    \end{subfigure}
    \hfill
    \begin{subfigure}[t]{0.45\textwidth}
        \centering
        \includegraphics[width=\textwidth]{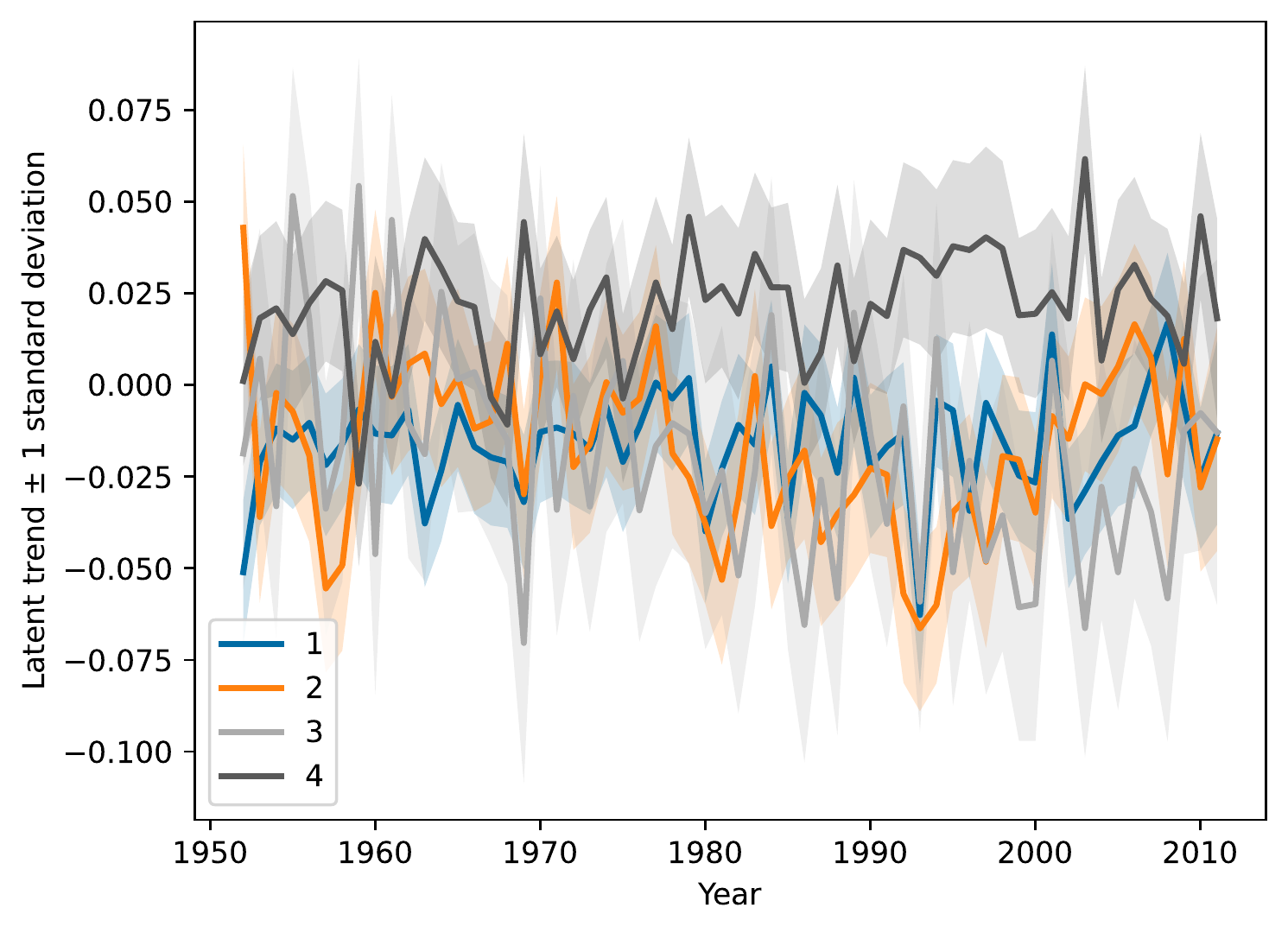}
        \caption{Trend of latent process. The shaded region is $\pm$ 1 standard deviation.}
    \end{subfigure}
    \hfill
    \begin{subfigure}[t]{0.45\textwidth}
        \centering
        \includegraphics[width=\textwidth]{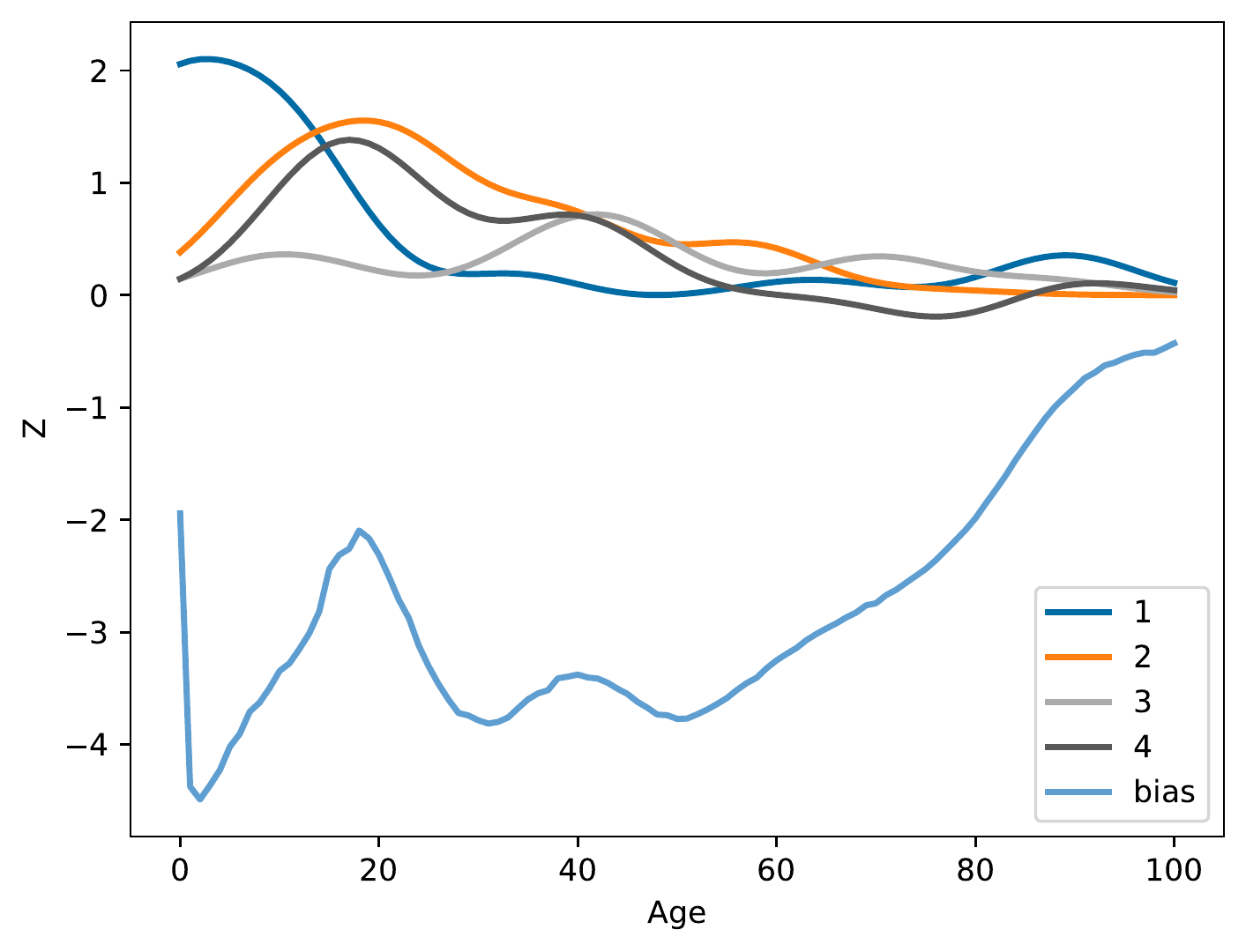}
        \caption{Factor loadings for the model with 4 latent dimensions}
        \label{fig:radial_loadings}
    \end{subfigure}
    \hfill
    \begin{subfigure}[t]{0.45\textwidth}
        \centering
        \includegraphics[width=\textwidth]{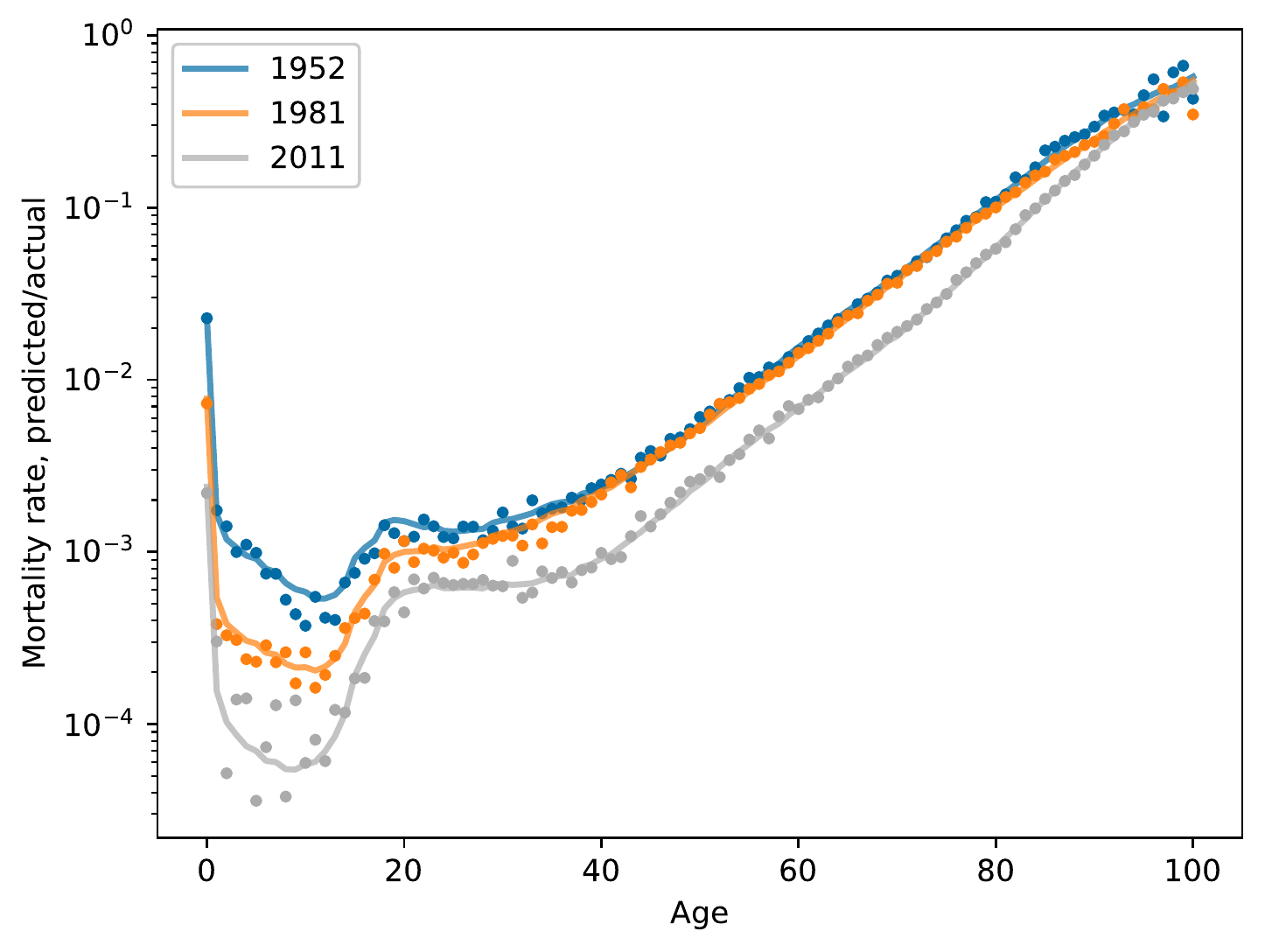}
        \caption{In-sample fit of model mortality rates. Dots are observed rates and solid lines are the fitted model. This model is fitted on data from 1952 to 2011.}
        \label{fig:radial_mortality_rates_in_sample_fit_sweden}
    \end{subfigure}
    \caption{Radial basis model: Illustration of the model fitted to Swedish data from 1952 to 2011 with 4 latent dimensions.}
    \label{fig:radial_model_illustration_sweden}
\end{figure}

\begin{figure}
    \centering
    \includegraphics[width=0.5\textwidth]{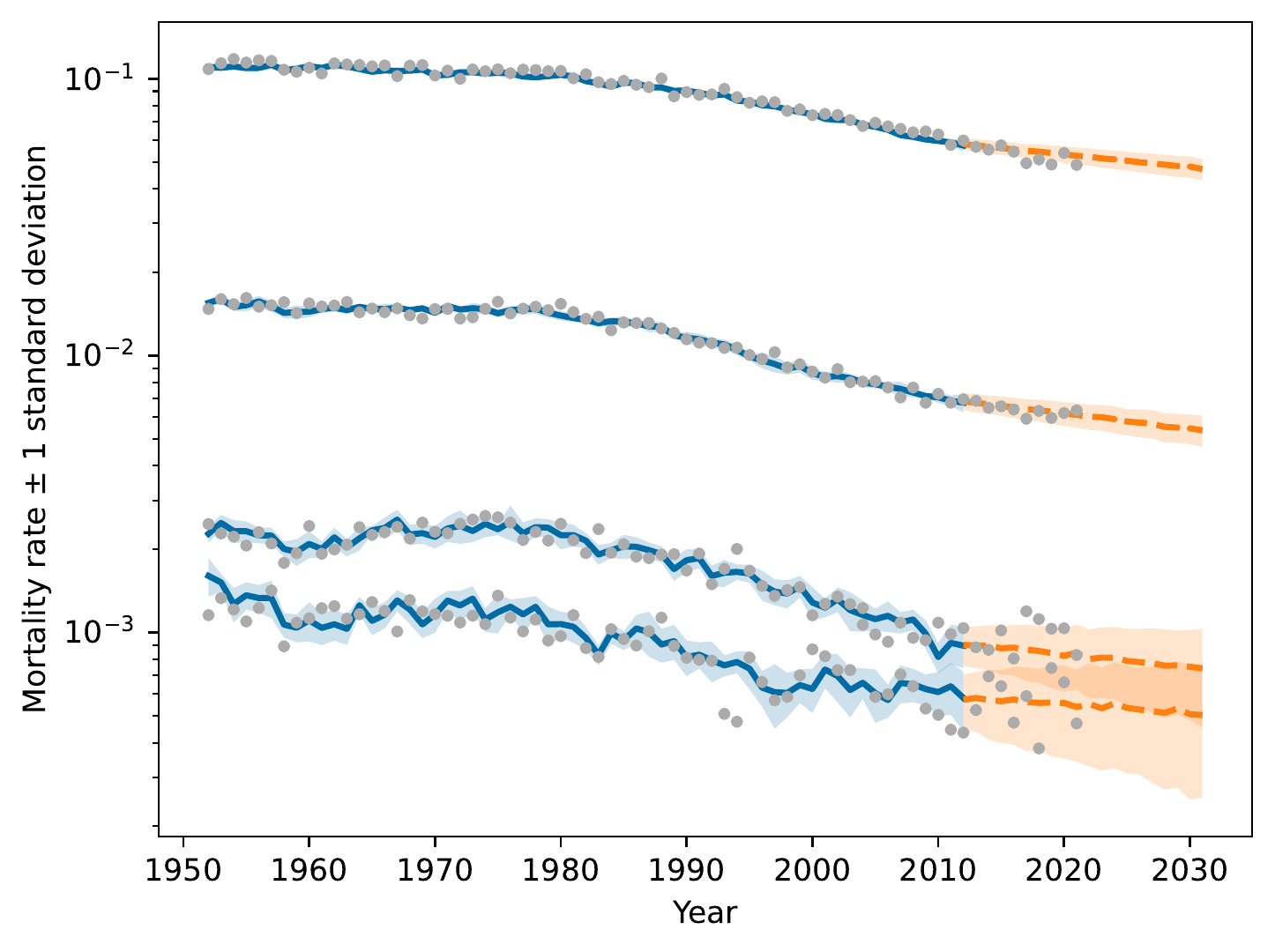}
    \caption{Radial basis model: Mortality forecasts for the age groups 20, 40, 60 and 80 (from the bottom up). The solid line is the smoothed mortality rate, the dashed line is the mean forecasted mortality rate and the shaded region is $\pm$ 1 standard deviation of the forecasted realised mortality rate. Dots indicate observed mortality rates.}\label{fig:radial_model_forecast_sweden}
\end{figure}

In Figure \ref{fig:radial_model_evaluation_sweden} we see that four latent dimensions and 15 radial basis functions give the best out-of-sample performance. In Figure \ref{fig:radial_model_illustration_sweden} the model fit is illustrated. In particular, we note in figures \ref{fig:radial_loadings} and \ref{fig:radial_mortality_rates_in_sample_fit_sweden} that the radial basis functions do indeed make both the factor loadings and the fitted mortality curves more smooth, compared to the affine model. In Figure \ref{fig:radial_model_forecast_sweden} we see that the forecasted mortality rates are quite similar to the affine model.

\subsection{Comparison}
In this section we compare our two models to two other commonly used mortality models, the Lee-Carter model with Poisson distribution from \cite{brouhns2002poisson} and the model from \cite{plat2009stochastic}. Both are fitted using the StMoMo package in \textsf{R} \citep{stmomo2018}. They both model mortality as
$$
D_{a,t}\sim \mathsf{Poisson}(E_{a,t}\exp\pa{\eta_{a,t}}),
$$
where in the Lee-Carter model,
$$
\eta_{a,t} = \alpha_a + \beta_a \kappa_t,
$$
where $\alpha_a$ and $\beta_a$ are factor loadings and $\kappa_t$ is the dynamic factor, modelled as a random walk with drift. In the Plat model,
$$
\eta_{a,t} = \alpha_a + \kappa_t^{(1)} + (\bar a - a)\kappa_t^{(2)} + (\bar a - a)^+\kappa_t^{(3)} + \gamma_{t-a}.
$$
Here the $\kappa_t^{(\cdot)}$s follow a multivariate random walk with drift and $\gamma_{t-a}$ is ARIMA(2,0,0) with intercept. The unknowns are estimated using maximum likelihood, and then the dynamic factors are modelled and forecasted. 

\begin{figure}
    \centering
    \begin{subfigure}[t]{0.49\textwidth}
        \centering
        \includegraphics[width=\textwidth]{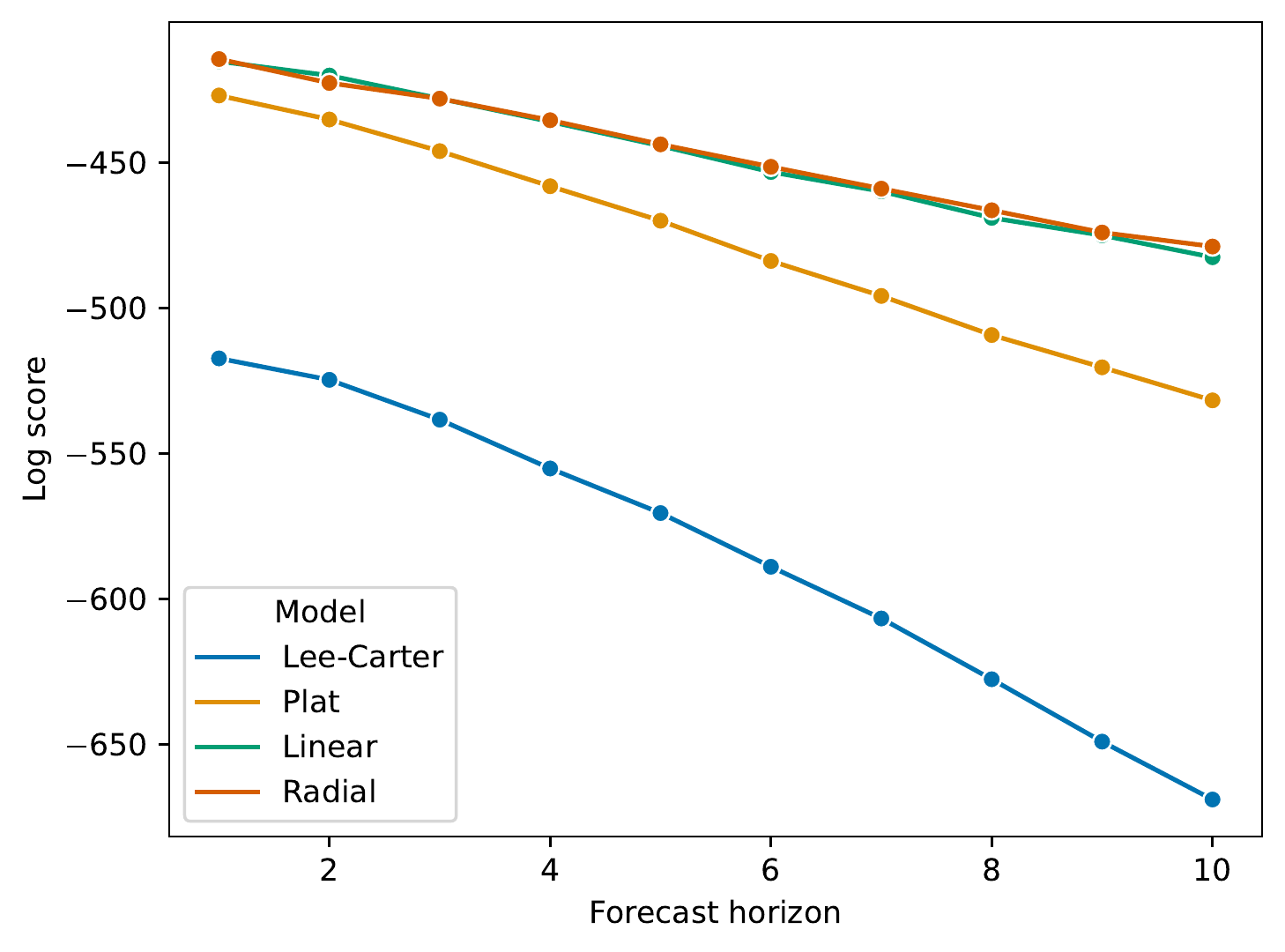}
        \caption{}
    \end{subfigure}
    \hfill
    \begin{subfigure}[t]{0.49\textwidth}
        \centering
        \includegraphics[width=\textwidth]{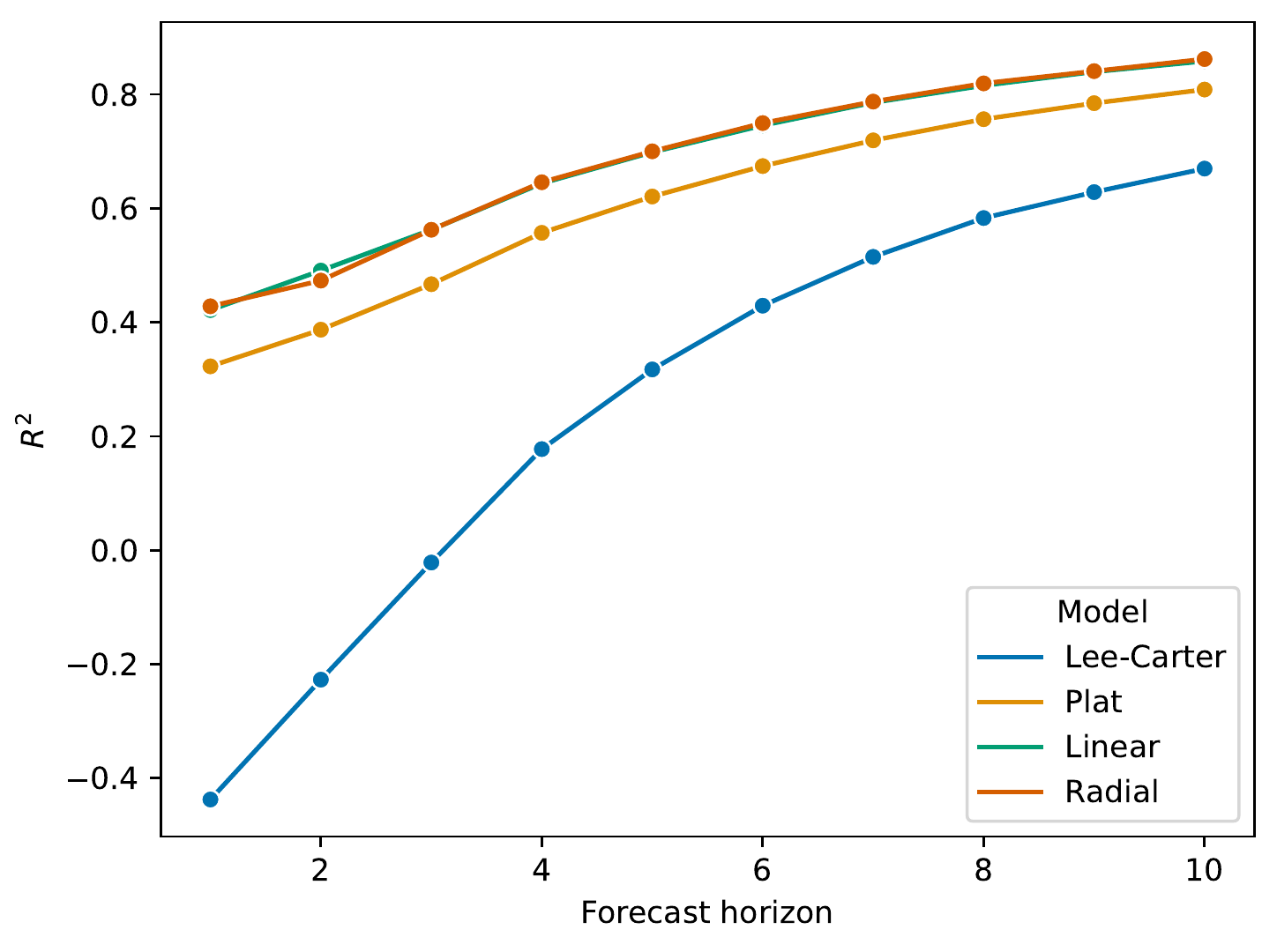}
        \caption{}
    \end{subfigure}
    \hfill
        \caption{Comparison of our model and the Lee-Carter and Plat models. Each model is fitted using 60 years of data and evaluated on the following 10 years. The figures show that both the affine and radial basis models perform well.}
        \label{fig:model_comparison}
\end{figure}

The model performance is summarized in Table \ref{table:model_comparision_sweden} and Figure \ref{fig:model_comparison} where the log-score and $R^2$ are shown for each model. We see that our two models perform almost identically and improve substantially on the two compared models.

\begin{table}[h!]
    \centering
    \begin{tabular}{l l l} 
        \hline
        Model & Log-score & $R^2$ \\  
        \hline\hline
        Affine & -448.2 & 0.686 \\
        Radial basis & -447.3 & 0.687 \\
        Lee-Carter & -584.6 & 0.263\\ 
        Plat & -477.6 & 0.610 \\
        \hline
    \end{tabular}
    \caption{Out-of-sample model evaluation metrics for Sweden. Numbers are averaged over all forecast horizons and rolling windows.}
    \label{table:model_comparision_sweden}
\end{table}
\newpage
\section{Conclusions}\label{sec:conclusions}
In this paper, we have considered a state-space model for mortality forecasting and we have shown how it is possible to fit such a model using variational inference. Using variational inference it is possible to not only use a Poisson likelihood for the observed number of deaths but also to estimate the complete model in one step. The model is also flexible in that, for example, we can consider different functions for projecting the latent variables to the mortality curve. We considered both affine functions and radial basis functions, but the practitioner has the freedom to choose other classes of functions without complication. Another advantage is that the model is implemented in Pyro, so that very little custom code is needed. Finally, we show that our model and inference method outperform other popular methods.

\section*{Data availability statement}
The data used in this paper can be downloaded free of charge from \url{https://www.mortality.org}.
    
\bibliographystyle{apalike}
\bibliography{mortality}

\end{document}